\input harvmac
\input amssym

\def\p{\partial}

\def\rt{\rightarrow}

\def\zb{\overline{z}}

\def\taub{\overline{\tau}}

\def\Ab{\overline{A}}
\def\Fb{\overline{F}}

\def\zb{\overline{z}}
\def\Lc{{\cal L}}
\def\Lcb{\overline{\cal L}}

\def\Wc{{\cal W}}
\def\Wcb{{\overline{\cal W}}}

\def\mub{\overline{\mu}}

\def\Wh{\hat{W}}
\def\Lh{\hat{L}}

\def\Ac{{\cal A}}
\def\Acb{ {\overline{\cal A}}}


\def\rs{r_{\star}}
\def\rts{\tilde{r}_{\star}}

\def\rar{\rightarrow}
\def\ml{{\cal L}}

\def\rti{\tilde{r}}
\def\vphi{\varphi}

\lref\WittenHC{
  E.~Witten,
  ``(2+1)-Dimensional Gravity as an Exactly Soluble System,''
  Nucl.\ Phys.\  B {\bf 311}, 46 (1988).
}

\lref\AchucarroVZ{
  A.~Achucarro and P.~K.~Townsend,
  ``A Chern-Simons Action for Three-Dimensional anti-De Sitter Supergravity
  Theories,''
  Phys.\ Lett.\  B {\bf 180}, 89 (1986).
}

\lref\BrownNW{
  J.~D.~Brown and M.~Henneaux,
  ``Central Charges in the Canonical Realization of Asymptotic Symmetries: An
  Example from Three-Dimensional Gravity,''
  Commun.\ Math.\ Phys.\  {\bf 104}, 207 (1986).
}


\lref\HaggiManiRU{
  P.~Haggi-Mani and B.~Sundborg,
  ``Free large N supersymmetric Yang-Mills theory as a string theory,''
  JHEP {\bf 0004}, 031 (2000)
  [arXiv:hep-th/0002189].
}

\lref\KonsteinBI{
  S.~E.~Konstein, M.~A.~Vasiliev and V.~N.~Zaikin,
  ``Conformal higher spin currents in any dimension and AdS/CFT
  correspondence,''
  JHEP {\bf 0012}, 018 (2000)
  [arXiv:hep-th/0010239].
}

\lref\SundborgWP{
  B.~Sundborg,
  ``Stringy gravity, interacting tensionless strings and massless higher
  spins,''
  Nucl.\ Phys.\ Proc.\ Suppl.\  {\bf 102}, 113 (2001)
  [arXiv:hep-th/0103247].
}

\lref\MikhailovBP{
  A.~Mikhailov,
  ``Notes on higher spin symmetries,''
  arXiv:hep-th/0201019.
}

\lref\SezginRT{
  E.~Sezgin and P.~Sundell,
  ``Massless higher spins and holography,''
  Nucl.\ Phys.\  B {\bf 644}, 303 (2002)
  [Erratum-ibid.\  B {\bf 660}, 403 (2003)]
  [arXiv:hep-th/0205131].
}

\lref\KlebanovJA{
  I.~R.~Klebanov and A.~M.~Polyakov,
  ``AdS dual of the critical O(N) vector model,''
  Phys.\ Lett.\  B {\bf 550}, 213 (2002)
  [arXiv:hep-th/0210114].
}

\lref\GiombiWH{
  S.~Giombi and X.~Yin,
  ``Higher Spin Gauge Theory and Holography: The Three-Point Functions,''
  JHEP {\bf 1009}, 115 (2010)
  [arXiv:0912.3462 [hep-th]].
}

\lref\GiombiVG{
  S.~Giombi and X.~Yin,
  ``Higher Spins in AdS and Twistorial Holography,''$\quad \quad$
  arXiv:1004.3736 [hep-th].
}

\lref\HenneauxXG{
  M.~Henneaux and S.~J.~Rey,
  ``Nonlinear W(infinity) Algebra as Asymptotic Symmetry of Three-Dimensional
  Higher Spin Anti-de Sitter Gravity,''
  JHEP {\bf 1012}, 007 (2010)
  [arXiv:1008.4579 [hep-th]].
}

\lref\CampoleoniZQ{
  A.~Campoleoni, S.~Fredenhagen, S.~Pfenninger and S.~Theisen,
  ``Asymptotic symmetries of three-dimensional gravity coupled to higher-spin
  fields,''
  JHEP {\bf 1011}, 007 (2010)
  [arXiv:1008.4744 [hep-th]].
}

\lref\GaberdielAR{
  M.~R.~Gaberdiel, R.~Gopakumar and A.~Saha,
  ``Quantum W-symmetry in AdS$_3$,''
  JHEP {\bf 1102}, 004 (2011)
  [arXiv:1009.6087 [hep-th]].
}

\lref\GaberdielPZ{
  M.~R.~Gaberdiel and R.~Gopakumar,
  ``An AdS$_3$ Dual for Minimal Model CFTs,''
  arXiv:1011.2986 [hep-th].
}

\lref\DouglasRC{
  M.~R.~Douglas, L.~Mazzucato and S.~S.~Razamat,
  ``Holographic dual of free field theory,''
  arXiv:1011.4926 [hep-th].
}

\lref\CastroCE{
  A.~Castro, A.~Lepage-Jutier and A.~Maloney,
  ``Higher Spin Theories in AdS$_3$ and a Gravitational Exclusion Principle,''
  JHEP {\bf 1101}, 142 (2011)
  [arXiv:1012.0598 [hep-th]].
}

\lref\GaberdielWB{
  M.~R.~Gaberdiel and T.~Hartman,
  ``Symmetries of Holographic Minimal Models,''
  arXiv:1101.2910 [hep-th].
}

\lref\FradkinKS{
  E.~S.~Fradkin and M.~A.~Vasiliev,
  ``On the Gravitational Interaction of Massless Higher Spin Fields,''
  Phys.\ Lett.\  B {\bf 189}, 89 (1987).
}

\lref\FradkinQY{
  E.~S.~Fradkin and M.~A.~Vasiliev,
  ``Cubic Interaction in Extended Theories of Massless Higher Spin Fields,''
  Nucl.\ Phys.\  B {\bf 291}, 141 (1987).
}


\lref\VasilievEN{
  M.~A.~Vasiliev,
  ``Consistent equation for interacting gauge fields of all spins in
  (3+1)-dimensions,''
  Phys.\ Lett.\  B {\bf 243}, 378 (1990).
}

\lref\VasilievCM{
  M.~A.~Vasiliev,
  ``Closed equations for interacting gauge fields of all spins,''
  JETP Lett.\  {\bf 51}, 503 (1990)
  [Pisma Zh.\ Eksp.\ Teor.\ Fiz.\  {\bf 51}, 446 (1990)].
}

\lref\VasilievAV{
  M.~A.~Vasiliev,
  ``More On Equations Of Motion For Interacting Massless Fields Of All Spins In
  (3+1)-Dimensions,''
  Phys.\ Lett.\  B {\bf 285}, 225 (1992).
}

\lref\BlencoweGJ{
  M.~P.~Blencowe,
  ``A Consistent Interacting Massless Higher Spin Field Theory In D = (2+1),''
  Class.\ Quant.\ Grav.\  {\bf 6}, 443 (1989).
}

\lref\BergshoeffNS{
  E.~Bergshoeff, M.~P.~Blencowe and K.~S.~Stelle,
  ``Area Preserving Diffeomorphisms And Higher Spin Algebra,''
  Commun.\ Math.\ Phys.\  {\bf 128}, 213 (1990).
}

\lref\ZamolodchikovWN{
  A.~B.~Zamolodchikov,
  ``Infinite Additional Symmetries In Two-Dimensional Conformal Quantum Field
  Theory,''
  Theor.\ Math.\ Phys.\  {\bf 65}, 1205 (1985)
  [Teor.\ Mat.\ Fiz.\  {\bf 65}, 347 (1985)].
}

\lref\KrausWN{
  P.~Kraus,
  ``Lectures on black holes and the AdS(3)/CFT(2) correspondence,''
  Lect.\ Notes Phys.\  {\bf 755}, 193 (2008)
  [arXiv:hep-th/0609074].
}

\lref\FronsdalRB{
  C.~Fronsdal,
  ``Massless Fields With Integer Spin,''
  Phys.\ Rev.\  D {\bf 18}, 3624 (1978).
}

\lref\DidenkoTD{
  V.~E.~Didenko and M.~A.~Vasiliev,
  ``Static BPS black hole in 4d higher-spin gauge theory,''
  Phys.\ Lett.\  B {\bf 682}, 305 (2009)
  [arXiv:hep-th/0906.3898].
}

\lref\KochCY{
  R.~d.~M.~Koch, A.~Jevicki, K.~Jin and J.~P.~Rodrigues,
  ``$AdS_4/CFT_3$ Construction from Collective Fields,''
  Phys.\ Rev.\  D {\bf 83}, 025006 (2011)
  [arXiv:1008.0633].
}


\lref\GK{
  M.~Gutperle and P.~Kraus,
  ``Higher Spin Black Holes,''
  JHEP {\bf 1105}, 022 (2011)
  [arXiv:1103.4304 [hep-th]].
}

\lref\AhnPV{
  C.~Ahn,
  ``The Large N 't Hooft Limit of Coset Minimal Models,''
  arXiv:1106.0351 [hep-th].
}

\lref\GaberdielZW{
  M.~R.~Gaberdiel, R.~Gopakumar, T.~Hartman and S.~Raju,
  ``Partition Functions of Holographic Minimal Models,''
  arXiv:1106.1897 [hep-th].
}

\lref\BershadskyBG{
  M.~Bershadsky,
  ``Conformal field theories via Hamiltonian reduction,''
  Commun.\ Math.\ Phys.\  {\bf 139}, 71 (1991).
}

\lref\PolyakovDM{
  A.~M.~Polyakov,
  ``Gauge Transformations and Diffeomorphisms,''
  Int.\ J.\ Mod.\ Phys.\  A {\bf 5}, 833 (1990).
}

\lref\BilalCF{
  A.~Bilal,
  ``W algebras from Chern-Simons theory,''
  Phys.\ Lett.\  B {\bf 267}, 487 (1991).
}

\lref\DynkinUM{
  E.~B.~Dynkin,
  ``Semisimple subalgebras of semisimple Lie algebras,''
  Trans.\ Am.\ Math.\ Soc.\  {\bf 6}, 111 (1957).
}

\lref\BaisBS{
  F.~A.~Bais, T.~Tjin and P.~van Driel,
  ``Covariantly coupled chiral algebras,''
  Nucl.\ Phys.\  B {\bf 357}, 632 (1991).
}

\lref\ForgacsAC{
  P.~Forgacs, A.~Wipf, J.~Balog, L.~Feher and L.~O'Raifeartaigh,
  ``Liouville and Toda Theories as Conformally Reduced WZNW Theories,''
  Phys.\ Lett.\  B {\bf 227}, 214 (1989).
}

\lref\deBoerIZ{
  J.~de Boer and T.~Tjin,
  ``The Relation between quantum W algebras and Lie algebras,''
  Commun.\ Math.\ Phys.\  {\bf 160}, 317 (1994)
  [arXiv:hep-th/9302006].
}

\lref\ChangMZ{
  C.~M.~Chang and X.~Yin,
  ``Higher Spin Gravity with Matter in $AdS_3$ and Its CFT Dual,''
  arXiv:1106.2580 [hep-th].
}

\lref\MaldacenaRE{
  J.~M.~Maldacena,
  ``The Large N limit of superconformal field theories and supergravity,''
  Adv.\ Theor.\ Math.\ Phys.\  {\bf 2}, 231 (1998)
  [Int.\ J.\ Theor.\ Phys.\  {\bf 38}, 1113 (1999)]
  [arXiv:hep-th/9711200].
}

\lref\WittenQJ{
  E.~Witten,
  ``Anti-de Sitter space and holography,''
  Adv.\ Theor.\ Math.\ Phys.\  {\bf 2}, 253 (1998)
  [arXiv:hep-th/9802150].
}

\lref\GubserBC{
  S.~S.~Gubser, I.~R.~Klebanov and A.~M.~Polyakov,
  ``Gauge theory correlators from noncritical string theory,''
  Phys.\ Lett.\  B {\bf 428}, 105 (1998)
  [arXiv:hep-th/9802109].
}

\lref\JevickiSS{
  A.~Jevicki, K.~Jin and Q.~Ye,
  ``Collective Dipole Model of AdS/CFT and Higher Spin Gravity,''
  arXiv:1106.3983 [hep-th].
}

\lref\WaldNT{
  R.~M.~Wald,
  ``Black hole entropy is the Noether charge,''
  Phys.\ Rev.\  D {\bf 48}, 3427 (1993)
  [arXiv:gr-qc/9307038].
}

\lref\ProkushkinBQ{
  S.~F.~Prokushkin and M.~A.~Vasiliev,
  ``Higher spin gauge interactions for massive matter fields in 3-D AdS
  space-time,''
  Nucl.\ Phys.\  B {\bf 545}, 385 (1999)
  [arXiv:hep-th/9806236].
}

\lref\ProkushkinVN{
  S.~Prokushkin and M.~A.~Vasiliev,
  ``3-d higher spin gauge theories with matter,''
  arXiv:hep-th/9812242.
}

\lref\Alej{
A. Castro, E. Hijano Cubelos, A. Lepage-Jutier  and A. Maloney,  To appear}


\lref\DidenkoZD{
  V.~E.~Didenko, A.~S.~Matveev and M.~A.~Vasiliev,
  ``BTZ Black Hole as Solution of 3-D Higher Spin Gauge Theory,''
  Theor.\ Math.\ Phys.\  {\bf 153}, 1487 (2007)
  [Teor.\ Mat.\ Fiz.\  {\bf 153}, 158 (2007)]
  [arXiv:hep-th/0612161].
}

\lref\VasilievAR{
  M.~A.~Vasiliev,
  ``Higher spin theories and Sp(2M) invariant space-time,''
  arXiv:hep-th/0301235.
}

\lref\VasilievDC{
  M.~A.~Vasiliev,
  ``Relativity, causality, locality, quantization and duality in the S(p)(2M)
  invariant generalized space-time,''
  arXiv:hep-th/0111119.
}


\lref\VasilievQH{
  M.~A.~Vasiliev,
  ``Quantization on sphere and high spin superalgebras,''
  JETP Lett.\  {\bf 50}, 374 (1989)
  [Pisma Zh.\ Eksp.\ Teor.\ Fiz.\  {\bf 50}, 344 (1989)].
}

\lref\VasilievRE{
  M.~A.~Vasiliev,
  ``Higher Spin Algebras and Quantization on the Sphere and Hyperboloid,''
  Int.\ J.\ Mod.\ Phys.\  A {\bf 6}, 1115 (1991).
}

\lref\BordemannZI{
  M.~Bordemann, J.~Hoppe and P.~Schaller,
  ``Infinite dimensional matrix algebras,''
  Phys.\ Lett.\  B {\bf 232}, 199 (1989).
}

\lref\FradkinQK{
  E.~S.~Fradkin and V.~Y.~Linetsky,
  ``Supersymmetric Racah basis, family of infinite dimensional superalgebras,
  SU(infinity + 1|infinity) and related 2-D models,''
  Mod.\ Phys.\ Lett.\  A {\bf 6}, 617 (1991).
}


\Title{\vbox{\baselineskip14pt
}} {\vbox{\centerline {Spacetime Geometry in Higher Spin Gravity}}}
\centerline{Martin Ammon,
  Michael Gutperle, Per
Kraus and Eric Perlmutter\foot{ammon@physics.ucla.edu, gutperle@ucla.edu, pkraus@ucla.edu, perl@physics.ucla.edu}}
\bigskip
\centerline{\it{Department of Physics and Astronomy}}
\centerline{${}$\it{University of California, Los Angeles, CA 90095,USA}}

\baselineskip14pt

\vskip .3in

\centerline{\bf Abstract}
\vskip.2cm

Higher spin gravity is an interesting toy model of stringy geometry.
Particularly intriguing is the presence of higher spin gauge transformations that redefine notions of invariance in gravity: the existence of event horizons and singularities in the metric become gauge dependent.  In previous work, solutions of spin-3 gravity in the SL(3,R) $\times$ SL(3,R) Chern-Simons formulation were found, and were proposed to play the role of black holes.  However, in the gauge employed there, the spacetime metric describes a traversable wormhole connecting two asymptotic regions, rather than a black hole.  In this paper, we show explicitly that under a higher spin gauge transformation these solutions can be transformed
to describe black holes with manifestly smooth event horizons, thereby changing the spacetime causal structure.
A related aspect is that the Chern-Simons theory admits two distinct AdS$_3$
vacua with different asymptotic ${W}$-algebra symmetries and central charges.  We show that these vacua are connected by an explicit, Lorentz symmetry-breaking RG flow, of which our solutions represent finite temperature generalizations. These features will be present in any SL(N,R) $\times$ SL(N,R) Chern-Simons theory of higher spins.

\Date{June  2011}
\baselineskip15pt


\newsec{Introduction}

The AdS/CFT correspondence \refs{\MaldacenaRE,\WittenQJ,\GubserBC} is a very powerful tool to understand aspects of both strongly coupled  gauge theories as well as gravitational theories. It is of great importance to find more examples  of the correspondence,  as well as to move away from the semiclassical gravity/large N - large 't Hooft coupling limit in which most work in AdS/CFT has been done.  It has long been speculated that the proper description of spacetime geometry at the string scale might involve a large enhanced gauge symmetry, corresponding to the infinite tower of string modes.
Recently,  theories with massless higher spin fields  \refs{\FronsdalRB,\FradkinKS,\FradkinQY,\VasilievEN, \VasilievCM,\VasilievAV} have been reevaluated in the context of the gauge/gravity correspondence \refs{\HaggiManiRU,\KonsteinBI,\SezginRT,\SundborgWP,\MikhailovBP}.

An interesting conjecture  \KlebanovJA\  proposes that  the  dual of the three dimensional O(N) vector model in the large $N$ limit   is a Vasiliev higher spin theory in AdS$_4$. In the last two years a substantial amount of evidence for this conjecture has  accumulated, see e.g.  \refs{
\GiombiWH,\GiombiVG,\KochCY,\DouglasRC,\JevickiSS}.

Higher spin theories in three dimensions \BlencoweGJ\ are considerably simpler than theories in four or more dimensions.  This is due to the fact that it is possible to truncate the infinite tower of higher spin fields to fields of spin $s\leq N$. Furthermore, the complicated nonlinear interactions of the higher spin fields can be reformulated in  terms of a Lagrangian  SL(N,R)$\times$ SL(N,R) Chern-Simons theory   \refs{\BergshoeffNS, \BordemannZI,\VasilievQH,\VasilievRE,\FradkinQK,\HenneauxXG,\CampoleoniZQ} generalizing the Chern-Simons formulation of three dimensional AdS gravity \refs{\AchucarroVZ,\WittenHC},  which is obtained as a special case  by setting  $N=2$.
For $N>2$ the underlying asymptotic Virasoro  symmetry is extended to a $W_N$ algebra \ZamolodchikovWN\
 via the (classical) Drinfeld-Sokolov reduction of the Chern-Simons theory \refs{\HenneauxXG,\CampoleoniZQ,\GaberdielAR}.

 In an interesting  paper  \GaberdielPZ\  a duality between three dimensional higher spin theories with $W_N$ symmetry and a $W_N$ minimal coset $SU(N)_k \otimes  SU(N)_1/ SU(N)_{k+1}$ CFT was conjectured, in a particular large $N$ 't Hooft limit.  This conjecture has been investigated further in \refs{ \GaberdielWB,\GaberdielZW,\ChangMZ,\AhnPV}.

The simplest higher spin theory is given by the $N=3$ Chern-Simons theory which describes the interactions of a graviton with a massless spin-3 field, and has an AdS$_3$ vacuum with two copies of the $W_3$ algebra as its asymptotic symmetry  group. The relation of the
Chern-Simons formulation and the higher spin theory was worked out in detail in \CampoleoniZQ\ and is briefly reviewed in section 2. An important feature of this theory is that gauge transformations of the  higher spin field  act nontrivially on the metric. As we
shall see, due to this property certain spacetime characteristics that we usually  think of as being invariantly defined --- such as the event horizon or curvature singularity of a black hole --- become gauge dependent.\foot{The fact that Riemannian geometry is not appropriate in higher spin theories is well-known, see e.g. \refs{\VasilievAR,\VasilievDC,\DidenkoTD}.}

Ordinary AdS$_3$ gravity contains BTZ black hole solutions. From the boundary CFT point of view these black holes carry nonzero charge with respect to the left- and right-
moving Virasoro zero modes, corresponding to  the mass and angular momentum of
the black hole\foot{For a review and references on $AdS_3/CFT_2$ and BTZ black
holes see \KrausWN. }. The $W_3$ algebra contains  additional left and right moving
spin-3 zero modes and it is natural to ask whether generalized black hole solutions exist that carry these charges (for work on black holes in higher spin theories from different perspectives see \refs{\CastroCE,\DidenkoTD}).

This question was answered in the affirmative in a recent paper by two of the present authors \GK, where a  solution carrying $W_3$ charge was found.
 For black holes in ordinary gravity (such as the BTZ black hole) the temperature of the
 black hole is determined by demanding that in   Euclidean signature solution the
 time circle closes off smoothly, and the entropy is given by the area of the horizon. In SL(3,R) gravity, due to the gauge dependence of the metric, both statements need to be replaced by a
 gauge invariant criterion. In \GK\ it was proposed that the holonomies of the Chern-Simons gauge fields around the Euclidean time circle should take the same values as for the BTZ black holes. It was shown that this criterion leads to a sensible thermodynamics obeying a first law; i.e.  the charges obey an integrability condition allowing one to integrate to find the thermodynamic partition function.  The entropy can then be calculated from the first law.

Curiously, for the most natural choice of  Chern-Simons gauge field, the resulting metric does not have a horizon at all, but is rather a traversable wormhole.  It was conjectured in \GK\  that a gauge transformation exists which transforms the wormhole solution to a black hole solution with a smooth horizon. In addition, it was suggested  that the conditions for such a gauge transformation to exist coincide with the holonomy conditions imposed in \GK. Some evidence for these conjectures was provided by employing perturbation theory in the higher spin charge.

The main result of the present paper is that we  show that the conjecture of   \GK\ and the interpretation of the black hole solution   is correct.  We find the gauge transformation that takes the wormhole to the black hole, and compute the explicit black hole metric, which is seen to be a spin-3 charged generalization of the BTZ solution.  We thus have a very concrete example of how two spacetime metrics with different causal structures can be gauge equivalent in higher spin gravity.

 A second important result of the paper is that apart from the AdS$_3$ vacuum which has $W_3$ symmetry,  there is a second AdS$_3$ vacuum which has a different asymptotic symmetry algebra, namely $W_3^{(2)}$, first constructed in \refs{\BershadskyBG,\PolyakovDM}. These two vacua correspond to the two inequivalent embeddings of SL(2,R) in SL(3,R).  It is simple to write down a solution representing a renormalization group flow from the $W_3^{(2)}$ CFT in the UV to the $W_3$ CFT in the IR.  It turns out that our black hole solution represents a finite temperature generalization of this RG flow solution.

The structure of the paper is as follows:
In section 2  we briefly review the Chern-Simons formulation of the spin-3 theory and  construct the AdS$_3$ vacuum which has  $W_3$ symmetry following \refs{\CampoleoniZQ\GK}. We also construct a second AdS$_3$ vacuum based on the non-principal embedding of SL(2,R) in SL(3,R) and show that the asymptotic symmetry algebra is $W_3^{(2)}$. Finally,  we construct  solutions which can be interpreted as renormalization group flows  between the two vacua. In section 3 we review the black hole solutions in ``wormhole" gauge found in \GK. In section 4  we construct the explicit gauge transformation taking the  solution from  ``wormhole" gauge to ``black hole" gauge, and establish that  the metric and spin-3 field  are smooth across the horizon precisely when the holonomy conditions are obeyed.  We close with a discussion of open questions and directions for future research. Some calculational details, as well as generalizations of some results to the case of general $N$, are relegated to appendices.

\noindent Note: We understand that work on related issues will appear in
\Alej.

\newsec{SL(3,R) Chern-Simons theory,  spin-3 gravity, and its AdS$_3$ vacua}

In this section we first quickly review the formulation of 3D higher spin
gravity in terms of SL(3,R) Chern-Simons theory.  We then discuss
the two distinct AdS$_3$ vacua of this theory, and their asymptotic symmetry algebras.   Finally, we exhibit an RG flow solution interpolating between these vacua, and study some of its properties.

\subsec{Chern-Simons action}

It was discovered   long  ago that Einstein gravity with a negative cosmological constant can be reformulated as a SL(2,R)$\times $ SL(2,R) Chern-Simons theory \refs{\AchucarroVZ,\WittenHC}.  It was shown in  \BergshoeffNS\ that a SL(N,R)$\times $ SL(N,R) Chern Simons theory corresponds to Einstein gravity coupled to $N-2$ symmetric tensor fields of spin $s=3,4, \ldots, N$.

In the following we will only consider the $N=3$ case, i.e., SL(3,R)$\times $ SL(3,R)
Chern-Simons theory, which corresponds to  spin-3 gravity in three dimensions with a negative cosmological constant.  Our conventions follow those in \GK.

 The action is
\eqn\abb{ S = S_{CS}[A] - S_{CS}[\Ab]}
where
\eqn\ac{  S_{CS}[A] ={k\over 4\pi} \int\! \Tr \left(A\wedge dA +{2\over 3} A\wedge A \wedge A\right)}
The 1-forms $A$ and $\Ab$ take values in the Lie algebra of SL(3,R). An explicit representation of the eight generators $L_i,~  i=-1,0,+1$ and $W_j,~ j=-2,-1,\cdots,+2$, as well as our conventions, is given in
appendix A.  The Chern-Simons level $k$ is related to the
Newton constant $G$ and AdS$_3$ radius $l$ as
\eqn\ad{k ={l\over 4G}}
We henceforth set $l=1$.

The Chern-Simons equations of motion correspond
to vanishing field strengths,
\eqn\ae{ F = dA + A\wedge A =0~,\quad \Fb= d\Ab+\Ab\wedge \Ab =0 }
To relate these to the spin-3 Einstein equations we introduce a vielbein
$e$ and spin connection $\omega$ as
\eqn\af{ A = \omega+e ~,\quad \Ab= \omega -e }
Expanding $e$ and $\omega$ in a basis of 1-forms $dx^\mu$,
the spacetime metric $g_{\mu\nu}$  and spin-3 field $\varphi_{\mu\nu\gamma}$
are identified as
\eqn\ag{ g_{\mu\nu} = {1\over 2} \Tr (e_\mu e_{\nu} )~,\quad \varphi _{\mu\nu\gamma} ={1\over 3! } \Tr ( e_{(\mu} e_\nu e_{\gamma)})}
where $\varphi_{\mu\nu\gamma}$ is totally symmetric as indicated.
Restricting to the SL(2,R) subalgebra generated by $L_i$, the flatness conditions \ae\
can be seen to be equivalent to Einstein's equations for the metric
$g_{\mu\nu}$ with a torsion free spin-connection.  More generally,
we find equations describing a consistent coupling of the metric to
the spin-3 field.

Acting on the metric and spin-3 field, the SL(3,R) $\times$ SL(3,R)  gauge  symmetries of the Chern-Simons theory turn into diffeomorphisms  along with
spin-3 gauge transformations (the Chern-Simons gauge transformation also include frame rotations, which leave the metric and spin-3 field invariant).    Under diffeomorphisms, the metric
and spin-3 field transform according to the usual tensor transformation rules.   The spin-3 gauge transformations are less familiar, as
they in general act nontrivially on both the metric and spin-3 field.
It is worth noting, though, that if we ignore the spin-3 gauge invariance,
then we can view the theory as a particular diffeomorphism invariant theory
of a metric and a rank-3 symmetric tensor field.

\subsec{The   $W_3$ AdS$_3$ vacuum}

We consider the following flat connections
\eqn\ai{\eqalign{ A_{AdS} & =  e^\rho L_1 dx^+ + L_0 d\rho \cr
\Ab_{AdS}  & =   -e^\rho L_{-1} dx^- - L_0 d\rho}}
where  $x^\pm = t \pm \phi$.  Using \ag, the corresponding metric and
spin-3 field are
\eqn\aia{\eqalign{ds^2 &=d\rho^2  -e^{2\rho} dx^+ dx^- \cr
\varphi_{\alpha\beta\gamma} &=0}}
The metric is that of an  AdS$_3$ of unit radius.

More generally, we can consider solutions that approach this vacuum
asymptotically, and work out  the resulting asymptotic symmetry algebra.  The analysis proceeds in a parallel fashion for the barred and unbarred connections, and so we'll just focus on the latter.
Also, since we will shortly present a more detailed computation for the second of our AdS$_3$ vacua, here we just sketch the steps.
Following \CampoleoniZQ, we consider the following form for the
connection
\eqn\aib{ A = \left(e^{\rho} L_1-{2\pi \over k} \Lc(x^+) e^{-\rho}L_{-1}-{\pi \over 2k}\Wc(x^+) e^{-2\rho} W_{-2}\right)dx^+ +L_0 d\rho }
In \CampoleoniZQ\ there is a parameter $\sigma$ accompanying the
W-generators that we are here setting to $\sigma =-1$.
Under gauge transformation,
\eqn\aic{ A \rt A +d\lambda +[A,\lambda] }
One then works out the most general $\lambda$ that preserves the form \aib.   Under these allowed gauge transformations the functions $\Lc(x^+)$ and $\Wc(x^+)$ transform.  The asymptotic symmetry algebra
is obtained from the Poisson brackets of the charges that generate
these transformations; see \CampoleoniZQ\ for details.  Alternatively (see section 4 of \GK\ for details), one can translate these variations into an operator product expansion for the symmetry currents, and the result is the $W_3$ algebra:
\eqn\aid{\eqalign{ T(z) T(0)& \sim  {3k \over z^4} + {2\over z^2}  T(0) + {1\over z} \p T(0)  \cr
T(z) \Wc(0) & \sim  {3\over z^2} \Wc(0) +{1\over z}\p \Wc(0) \cr
\Wc(z) \Wc(0) & \sim
 -{5  k \over \pi^2} {1\over z^6}+{10 \over \pi} {1\over z^4}\Lc(0)  +{5 \over \pi}{1\over z^3}\p \Lc(0) +{3 \over 2\pi}{1\over z^2} \p^2\Lc(0) -{1 \over 3\pi} {1\over z} \p^3 \Lc(0) \cr
&\quad -{32  \over 3k}{1\over z}\Lc(0)  \p \Lc(0) -{32  \over 3k}{1\over z^2} \Lc(0)^2  }}
Here we are using complex coordinates:  $z= x^+$.   The $T\Wc$ OPE
identifies $\Wc$ as a spin-3 current, i.e., as a dimension $(3,0)$ primary operator.    The same analysis for the barred connection gives rise to
anti-holomorphic $W_3$ algebra with a dimension $(0,3)$ current.

From the $TT$ OPE the central charge is
found to be
\eqn\aie{ c =6k}
which, using \ad, agrees with the usual Brown-Henneaux formula $ c = 3l/(2G)$ \BrownNW.

\subsec{The $W_3^{(2)}$  AdS$_3$ vacuum }

The procedure reviewed above for extracting the asymptotic symmetry
algebra is an example of the classical Drinfeld-Sokolov procedure, with which one can construct a $W$-algebra from an embedding of the SL(2,R) algebra inside a Lie algebra.
In the example just given,
the SL(2,R) algebra  was generated by $(L_1, L_0, L_{-1})$.
This is the
``principal embedding" in SL(3,R). It is characterized by the fact that the five $W$ generators transform as a spin 2 multiplet with respect to the SL(2,R) algebra.  As discussed, this embedding gives
rise to the $W_3$ algebra.

In the case of SL(3,R)  there is exactly one additional inequivalent  embedding of SL(2,R), up to conjugation\foot{Some relevant facts about SL(2,R) embeddings in SL(N,R) are reviewed in appendix E.}.
We define rescaled versions of $(W_2, L_0, W_{-2})$ so that they obey
the same commutation relations as $(L_1, L_0, L_{-1})$,
\eqn\aif{ \Wh_2 = {1\over 4}W_2,~\quad \Lh_0 = {1\over 2}L_0~,\quad
\Wh_{-2} =-{1\over 4} W_{-2} }
These have traces
\eqn\aig{\eqalign{ \Tr (\Lh_0 \Lh_0)& ={1\over 2}={1\over 4} \Tr (L_0 L_0)~,\cr
 \Tr(\Wh_2 \Wh_{-2} ) &= -1={1\over 4}\Tr(L_1 L_{-1}) }}
Note that the branching of the adjoint representation of SL(3,R) into SL(2,R) representations is different in this case. There is one spin-0 multiplet, given by $W_0$, and two spin-1/2 multiplets,  given by $(W_1,L_{-1})$ and $(L_{1},W_{-1})$. (This is in addition to a spin-1 multiplet given by the SL(2,R) generators themselves).

This embedding gives rise to another $W$-algebra known as $W_3^{(2)}$, sometimes  referred to as the Polyakov-Bershadsky algebra \refs{\BershadskyBG,\PolyakovDM}.   Next we
work out the corresponding AdS$_3$ vacuum and its asymptotic $W_3^{(2)}$ symmetry algebra. Its central charge, which we denote by $\hat{c}$, will soon be related to the central charge $c$ of the $W_3$ vacuum.

The unbarred gauge field includes fields for the highest weight generator in each multiplet
\eqn\wtta{A= e^{-\rho \hat L_0}  \Big( \hat W_2  - {\cal T} \hat W_{-2} + jW_0 + g_1L_{-1} + g_2W_{-1}\Big)  e^{\rho \hat L_0} dx^+ + \hat L_0 d\rho}
We parameterize a general gauge transformation as follows
\eqn\wttb{\eqalign{\lambda &= e^{-\rho \hat L_0} \Big( \epsilon_1 \hat W_2+ \epsilon_0  \hat L_0 + \epsilon_{-1}\hat W_{-2}  +  \gamma W_0 + \delta_{1} L_1 + \delta_{-1}W_{-1}   + \rho_{1}  W_1 + \rho_{-1}L_{-1} \Big) e^{\rho \hat L_0}  }}
Here the $x^+$ dependence of the fields and the transformation parameters is suppressed. We demand that the gauge transformation
\eqn\wwtc{ \delta A = d\lambda +[A,\lambda] }
respect the form of \wtta.  The conditions determine the transformations of the fields $(T,j,g_{1}, g_{2})$ dependent on the parameters $(\epsilon_1, \gamma, \delta_1, \rho_1)$.
Defining new fields
\eqn\wwtd{\eqalign{ j&= {9\over 2 \hat{c}} U\cr
{\cal T} &= - {6\over \hat{c}} T - {27 \over \hat{c}^2} U^2\cr
g_1 &= {3\over \sqrt{2} \hat{c}}\big( G_+ +G_-\big) \cr
g_2 &= {3\over \sqrt{2} \hat{c}}\big( G_+ -G_-\big)  }}
as well as transformation parameters
\eqn\wwte{\eqalign{\epsilon_1&=\epsilon\cr
\gamma&=- {1\over 2}\eta + {9 \over 2 \hat{c}} U \epsilon \cr
\delta_{1}&= {1\over 2 \sqrt{2}} (\alpha_+ +\alpha_-) \cr
\rho_{1} &= {1\over 2 \sqrt{2}} (\alpha_+ -\alpha_-) \cr
}}
the transformation rules become
\eqn\wttf{\eqalign{\delta U&=  \epsilon' U + \epsilon U' - \alpha_+ G_+ +\alpha_- G_- - {\hat{c}\over 9} \eta'\cr
\delta T &= {\hat{c}\over 12} \epsilon'''+ 2 \epsilon' T + \epsilon T' + {3\over 2} \alpha'_+ G_+ +{1\over 2} \alpha_+ G_+' + {3\over 2} \alpha'_- G_- +{1\over 2} \alpha_- G_-' + \eta' U\cr
\delta G_+ &= {\hat{c}\over 6} \alpha_-'' + {3\over 2} \epsilon' G_+ + \epsilon G_+'  + \alpha_- (T +{18\over \hat{c}} U^2  + {3\over 2}  U') + 3 \alpha_-' U + \eta G_+\cr
\delta G_- &= -{\hat{c}\over 6} \alpha_+'' + {3\over 2} \epsilon' G_- + \epsilon G_-'  - \alpha_+ (T +{18\over \hat{c}} U^2  - {3\over 2}  U') + 3 \alpha_+' U - \eta G_-}}
where the prime denotes a derivative with respect to $x^+$.
These are identical to those  generated according to
\eqn\wwth{\delta {\cal O } = 2\pi {\rm Res} \Big( J(z) O(0) \Big) }
with the current   (denoting the operators by the same symbols in an abuse of notation)
\eqn\wwtj{ J= {1\over 2\pi} \Big( \epsilon T + \eta U +\alpha_+ G_+ +\alpha_- G_-\Big)}
 and the operator product expansion
\eqn\wttg{\eqalign{ U(z)U(0) &\sim  -{\hat{c}\over 9z^2 }\cr
U(z) G_{\pm}(0) &\sim \pm {1\over z} G_{\pm}(0) \cr
T(z) U(0) &\sim {1\over z^2} U(0) + {1\over z} \partial U(0)  \cr
T(z) G_\pm(0) &\sim {3\over 2z^2} G_\pm(0) + {1\over z} \partial G_\pm(0)  \cr
T(z) T(0) &\sim {\hat{c}\over 2z^4}+ {2\over z^2} T(0) + {1\over z} \partial T(0)   \cr
G_+(z) G_-(0) &\sim -{\hat{c}\over 3z^3} +{3\over z^2} U(0) -{1\over z}\big( T(0) + {18\over \hat{c}} U(0)^2-{3\over 2} \partial U(0) \big) }}
This is the  classical $W_3^{(2)}$ algebra (see e.g. \BilalCF). Apart from the stress energy tensor $T$, we have weight $3/2$ primaries, $G_\pm$, as well as a weight one current, $U$.  Note that the conformal weight of each field is obtained from   the SL(2,R) spin by adding one. This is a general feature as discussed in appendix E.  Due to the presence of the two spin-3/2 bosonic currents, the $W_3^{(2)}$ algebra can be thought of as a sort of bosonic analog of the ${\cal N}=2$ superconformal algebra.  Unlike the latter, the
$W_3^{(2)}$ is nonlinear, as seen by the appearance of $U^2$ in the $G_+ G_-$ OPE.

The AdS vacuum for the SL(2,R) embedding \aif\  is given by gauge connections
\eqn\aih{\eqalign{ A_{AdS} & =  e^\rho \Wh_2 dx^+ + \Lh_0 d\rho \cr
\Ab_{AdS}  & =   -e^\rho \Wh_{-2} dx^- - \Lh_0 d\rho}}
which yields
\eqn\aii{\eqalign{ds^2 &={1\over 4}\left(d\rho^2  -e^{2\rho} dx^+ dx^-\right)  \cr
\varphi_{\alpha\beta\gamma} &=0}}
The metric now describes an AdS$_3$ of radius $1/2$.

Note that the central charge $\hat{c}$ was not determined by the  procedure leading to \wttg.  This is a consequence of the fact that the Chern Simons level $k$ does not enter in \wwtc.
The central charge can, however, be determined by the following argument.

Comparing the  two SL(2,R) embeddings,  the generators that appear obey the same SL(2,R)  commutation relations.  The only difference arises from
the rescaled trace relations \aig.   This has the effect of reducing
the overall normalization of the Chern-Simons action restricted to the
SL(2,R) subalgebra by a factor $1/4$ compared to before,  which is to say that $k$ is
effectively replaced by $k/4$.  The central charge of the $W_3^{(2)}$ vacuum is therefore modified:
\eqn\aik{ \hat{c} =  {1\over 4} c = {3k\over 2}}

This result can also be established from the metric point of view.  If instead of using \ag\ to define the metric
we use $\hat{g}_{\mu\nu} = 2 \Tr (e_\mu e_\nu)$, then the
$W_3^{(2)}$ AdS$_3$ vacuum  will again have unit radius.  The action expressed in
terms of the hatted metric will take the same form as in the unhatted
case, except for an overall factor of $1/4$ from the rescaled trace.
Applying the Brown-Henneaux formula then again yields \aik.

Yet another way to arrive at  this result is to compute the Poisson brackets of the charges generating the asymptotic symmetry transformations.  The rescaled trace relations just lead to a factor in front of the action that
leads to rescaled versions of the canonical momenta.  Once again, the effect is just to replace $k$ by $k/4$.

Given  the metric \aii\ with radius $1/2$, one might be tempted to conclude that $\hat{c} = {1\over 2}c$ by applying the Brown-Henneaux formula directly.   However, a proper analysis has to take into account both
the scale size of the AdS$_3$ and the effective Newton constant, which
this argument does not do.

\subsec{RG flow between $W_3^{(2)}$ and $W_3$ vacua}

It is very easy to write down a solution that interpolates between
the $W_3^{(2)}$ vacuum in the UV and the $W_3$ vacuum in the IR. We take
\eqn\aja{\eqalign{   A & =  \lambda e^{\rho}L_1 dx^+
 + e^{2\rho} \Wh_2 dx^- +L_0 d\rho \cr
\Ab & = -\lambda e^{\rho} L_{-1} dx^- - e^{2\rho} \Wh_{-2} dx^+ - L_0 d\rho  }}
Note that we have chosen to accompany $d\rho$ by $L_0$ rather than
$\Lh_0$, which accounts for the $e^{2\rho}$ factors multiplying the
W-generators.  Also,  for reasons to be explained momentarily, compared to \aih\ we take the $\Wh_2$ term in $A$ to be associated with $dx^-$ rather than $dx^+$.  The parameter $\lambda$ is arbitrary, and can
be set to any desired value by shifting $\rho$ and scaling $x^\pm$.
At large $\rho$ the connections approach \aih\ (after rescaling $\rho$ and exchanging $x^+ \leftrightarrow x^-$), while at small $\rho$ they approach \ai.

The corresponding metric and spin-3 fields are
\eqn\ajb{\eqalign{ ds^2& = d\rho^2 - \left({1\over 4} e^{4\rho}+\lambda^2 e^{2\rho}\right) dx^+ dx^- \cr
\varphi_{\alpha\beta\gamma}dx^\alpha dx^\beta dx^\gamma  & = {1\over 3!} \Tr ( e e e) = -{1\over 8} \lambda^2 e^{4\rho} (dx^+)^3 + {1\over 8} \lambda^2 e^{4\rho} (dx^-)^3 }}
The metric interpolates between the two AdS$_3$ vacua at large and small $\rho$.    In an orthonormal frame, the spin-3 field vanishes
asymptotically at large and small $\rho$, but is nonzero in between.
Additionally, while the metric preserves Lorentz invariance in $x^\pm$, the
spin-3 field does not.

We now discuss the CFT interpretation of the RG flow from the standpoints of the UV and IR CFTs.   From the standpoint of the UV CFT with $W_3^{(2)}$ symmetry, the RG flow is triggered by the $\lambda$-terms in \aja. In the UV
CFT $\lambda$ is a source for the spin-3/2 operators.\foot{It was in order to have this interpretation that we associated the $L_1$ and $\Wh_2$ generators in \aja\ with opposite chiralities.} This can be seen from the field equations, or by noting that under the UV scale invariance $\lambda$ has dimension $1/2=2-3/2$, which is correct for a source conjugate to a dimension $3/2$ operator.    Thus, the RG flow is initiated by adding to the UV CFT Lagrangian a relevant operator of dimension $3/2$.

Thinking in terms of the IR CFT with $W_3$ symmetry, the flow is initiated by adding the $\Wh_{\pm 2}$ terms, which correspond to adding to the Lagrangian the spin-3 currents.   These are irrelevant dimension 3 operators, and so deform the theory in the UV. They drive the theory
to the new UV fixed point with $W_3^{(2)}$ symmetry.

One surprising feature concerns the central charges.  We have
$c_{UV} = {3k\over 2}$ and $c_{IR} = 6k$,  and so $c_{IR} > c_{UV}$,
which seems at first to be in conflict with the c-theorem.  However, there is no actual contradiction since the proof of the c-theorem applies to Lorentz invariant RG flows (rotationally invariant in Euclidean signature), whereas here we are adding non-Lorentz invariant operators.   The fixed points of the RG flow are Lorentz invariant field theories, but the full flow is not.    While there is thus no
immediate conflict with the c-theorem, this result is still somewhat puzzling as it seems at odds with the usual intuition regarding the decrease of
degrees of freedom under RG flow.   The resolution of this puzzle may involve the fact that the UV CFT really has a family of stress tensors due to the presence of a U(1) current algebra.

While we leave a complete analysis of this and related RG flows to future work, it is useful to present a few results to aid in interpretation.   First, we consider linearized perturbation theory around the RG flow background.  This will allows us to determine the relation between UV and IR operators.  Focussing just on the $A$-connection, we turn on terms corresponding to nonzero currents in the UV and IR.  We thus add to \aja\ the terms
\eqn\ajc{\eqalign{ \delta A &= \left(T_{IR}e^{-\rho} L_{-1} + \Wc_{IR} e^{-2\rho} W_{-2} \right)dx^+  \cr
& + \left( J_{UV} W_0 + G^{(1)}_{UV}e^{-\rho}L_{-1} +  G^{(2)}_{UV}e^{-\rho}W_{-1} + T_{UV}e^{-2\rho}W_{-2}\right)dx^- }}
where all coefficients, $T_{IR}$ etc, are arbitrary functions of $(x^+,x^-)$.  Up to normalization, these coefficient functions are the
respective currents in the UV/IR.

Solving the linearized field equations, we find
\eqn\ajd{\eqalign{ J_{UV}&= {1\over 2 \lambda} T_{IR} \cr
G^{(1)}_{UV} & = -{2\over \lambda}\Wc_{IR} \cr
G^{(2)}_{UV} &= -{1\over 6\lambda^2} \p_+ T_{IR} \cr
T_{UV}&= {1\over 24\lambda^3} \p_+^2 T_{IR}        }}
along with
\eqn\aje{\eqalign{ \p_- T_{IR} &= -{2\over \lambda }\p_+ \Wc_{IR} \cr
\p_- \Wc_{IR} &= {1\over 24\lambda^3} \p_+^3 T_{IR} }}
From the first two relations in \ajd\ we see that the IR currents
$(T_{IR}, \Wc_{IR})$ are locally related to the UV currents $(J_{UV}, G^{(1)}_{UV})$.    In particular, the IR stress tensor is locally related to the UV spin-1 current, {\it not} to the UV stress tensor.   The
equations in \aje\ show that deep in the IR, which corresponds to
large $\lambda$,  the currents $(T_{IR}, \Wc_{IR})$ obey chiral conservation equations as expected.

The relation between $T_{UV}$ and $T_{IR}$ indicates that $T_{UV}$,
which is of course a dimension $2$ operator in the UV, acquires dimension $4$ in the IR.  This can be shown in more detail by computing the AdS/CFT two-point correlator $\langle T_{UV} T_{UV}\rangle$.
This computation is carried out in appendix B, and the result is, in momentum space and up to overall  normalization:
\eqn\ajf{ \langle T_{UV}(p) T_{UV}(-p) \rangle = {p_+^3 p_- \over \lambda^4 -{4\over 3}{p_+^4\over p_-^2}} }
At short distances we have ${p_+^4\over p_-^2}\gg \lambda^4$ and
so the UV result is
\eqn\ajg{{\rm UV}:\quad \langle T_{UV}(x) T_{UV}(0) \rangle \sim
\int\! {d^2p \over (2\pi)^2} {p_-^3 \over p_+} e^{ip\cdot x}\sim { 1\over (x^-)^4 }}
At long distances we take $ {p_+^4\over p_-^2} \ll \lambda^4$, and
expand \ajf\ to first subleading order, since the leading order term
is polynomial in momentum space and hence a contact term in position space.  The leading behavior is thus
\eqn\ajh{{\rm IR}:\quad \langle T_{UV}(x) T_{UV}(0) \rangle \sim
\int\! {d^2p \over (2\pi)^2} {p_+^7 \over p_-}e^{ip\cdot x} \sim { 1\over (x^+)^8 }}
This result shows that $T_{UV}$ goes from being a dimension $(2,0)$ operator in the UV, to being a dimension $(0,4)$ operator in the IR.  The fact that the operator goes from being rightmoving to leftmoving  is of course a consequence of the non-Lorentz invariant character of the RG flow.

\newsec{Review of spin-3 blackhole solutions in wormhole gauge}

In \GK\ the following solution was proposed to represent black holes carrying spin-3 charge:
\eqn\caa{\eqalign{ \Ac&= \Big(e^\rho L_1 -{2\pi \over k}\Lc  e^{-\rho}L_{-1} -{\pi \over 2k} \Wc e^{-2\rho} W_{-2} \Big)dx^+  \cr
& \quad +\mu \Big(e^{2\rho} W_2 -{4\pi \Lc \over k}W_0 +{4\pi^2 \Lc^2 \over k^2} e^{-2\rho} W_{-2} +{4\pi \Wc \over k}e^{-\rho} L_{-1}\Big)dx^-  +L_0 d\rho  \cr
 \Acb&= -\Big(e^\rho L_{-1} -{2\pi \over k}\Lcb e^{-\rho} L_{1} -{\pi \over 2k} \Wcb e^{-2\rho} W_{2} \Big)dx^- \cr
 & \quad -\mub \Big( e^{2\rho}  W_{-2} -{4\pi \Lcb \over k}W_0 +{4\pi^2 \Lcb^2 \over k^2} e^{-2\rho} W_{2} +{4\pi \Wcb \over k}e^{-\rho} L_{1}\Big) dx^+ -L_0 d\rho  }}
The structure of this solution is easy to understand. Focus on the $\Ac$-connection. As in \aib, to add energy and charge density to the
$W_3$ vacuum we should add to $\Ac_+$ terms involving $L_{-1}$ and $W_{-2}$, as seen in the top line of \caa.   For black holes, which represent states of thermodynamic
equilibrium, the energy and charge should be accompanied by their
conjugate thermodynamic potentials, which are temperature and spin-3 chemical potential.  We incorporate the former via the periodicity of
imaginary time, while the latter was shown in \GK\ by a Ward identity analysis to correspond to a
$\mu W_2$ term in $\Ac_-$.  The remaining $\Ac_-$ terms appearing in \caa\ are then fixed by the equations of motion.

We will henceforth restrict attention to the nonrotating case:
\eqn\cab{ \Lcb = \Lc~,\quad \Wcb=-\Wc~,\quad \mub = -\mu}
If we set $\mu = \Wc=0$, then the connections become those corresponding to a BTZ black hole asymptotic to the $W_3$ vacuum.
From the standpoint of this CFT, the $\mu e^{2\rho} W_2 dx^-$ term
represents a chemical potential for spin-3 charge, and the
$\Wc e^{-2\rho} W_{-2} dx^+$ term gives the value of the spin-3 charge. This solution is therefore interpreted as a generalization of the BTZ black hole to include nonzero spin-3 charge and chemical potential.

Another useful special case to consider is $\Lc = \Wc=0$ with $\mu \neq 0$.  After shifting $\rho$, the resulting connection is identified with the RG flow solution \aja\ with $\lambda = {1\over 2\sqrt{\mu}}$.
So for small $(\Lc, \Wc)$ and finite $\mu$,  we can think of this solution as representing an excited version of the RG flow.

The general non-rotating solution \caa-\cab, as written, can be thought of as depending on four free parameters:  three of these are  $(\Lc, \Wc, \mu)$, and the fourth is the inverse  temperature $\beta$, corresponding to the periodicity of imaginary time, $t \cong t+i\beta$.  However, we expect that there should only be a two-parameter family of physically admissible solutions: once one has specified the temperature and chemical potential the values of the energy and charge should be determined thermodynamically.  For the uncharged BTZ solution the relation between the energy and the temperature is obtained by demanding the absence of a conical singularity at the horizon in Euclidean signature.  The analogous procedure in the presence of spin-3 charge is more subtle, as was discussed in detail in \GK.   It was proposed there that one should compute the holonomy  of the Chern-Simons connection around the Euclidean time circle, and demand that its eigenvalues take the  fixed values $(0, 2\pi i, -2\pi i)$.  This was proposed as the gauge invariant characterization of the condition for the Euclidean time circle to smoothly close off at the horizon.  This condition has the virtue of being gauge invariant, reproducing known BTZ results in the uncharged limit, and, as shown in \GK, being compatible with the first law of thermodynamics.  In the next section we will find strong independent evidence for the validity of this proposal.

Writing
\eqn\cad{\tau = {i\beta \over 2\pi} }
corresponding to the modular parameter of the torus, the holonomy around the time circle is\foot{For completeness, in appendix C we provide the holonomy along the angular direction.}
\eqn\cae{ \omega = 2\pi (\tau \Ac_+ - \taub \Ac_-)}
The conditions on its eigenvalues can be recast as
\eqn\caf{ \det (\omega)=0~,\quad \Tr(\omega^2)+8\pi^2 =0 }
which, for the connection \caa, become explicitly
\eqn\cag{\eqalign{ 0&= -2048 \pi^2 \mu^3  \Lc^3+576 \pi  k  \mu \Lc^2 -864 \pi k \mu^2  \Wc \Lc +864 \pi  k \mu^3  \Wc^2-27 k^2 \Wc \cr
0 & =256 \pi^2  \mu^2 \Lc^2  +24 \pi k  \Lc-72 \pi k \mu   \Wc +{3k^2\over \tau^2}  }}

These relations pass an important consistency check.   We want to think of the black hole as a saddle point contribution to a partition function of the form $Z(\tau,\alpha)= \Tr [e^{8\pi^2 i (\tau \Lc + \alpha \Wc)}]$. But   since this implies $\Lc \sim \p Z /\p \tau$ and $\Wc\sim \p Z / \p \alpha$, $Z$ will exist only if the charge assignments following from \cag\ obey the integrability condition ${\p \Lc \over \p \alpha} = {\p \Wc \over \p \tau}$.  The consistency check is that upon taking $\alpha = \beta \mu$, one indeed finds that the conditions \cag\ imply integrability.  This is equivalent to saying that we can define
an entropy that is consistent with the first law of thermodynamics.
In contrast to ordinary gravity where we can use either the area law
or the value of the Euclidean action to  compute the entropy, neither
of these approaches are immediately applicable in the higher spin case, and so we need to proceed as described here. Note also that the relation $\alpha =\beta \mu$ should be thought of as being determined self-consistently together with the charge assignments.

Now that we have reduced the black hole solutions to a two-parameter family, we can work out the corresponding metric, which leads to a surprise.  Using \ag, we find a metric of the form (see \GK\ for the explicit formulas)
\eqn\cah{ ds^2 = d\rho^2 -{\cal F}(\rho)dt^2 + {\cal G}(\rho)d\phi^2 }
The metric reduces to BTZ in the uncharged limit, but
the surprise is that for any nonzero charge, ${\cal F}(\rho)$ and ${\cal G}(\rho)$ are both positive definite quantities. In particular, since ${\cal F}$ never vanishes there is no event horizon --- this geometry possesses a globally defined timelike Killing vector.  At large positive and negative $\rho$, both ${\cal F}$ and ${\cal G}$ have leading behavior $e^{4|\rho|}$,
corresponding to an AdS$_3$ metric of radius $1/2$, and so the metric
\cah\ describes a traversible wormhole connecting two asymptotic AdS$_3$ geometries. We recognize these asymptotic AdS$_3$ regions as the $W_3^{(2)}$ vacua discussed previously.   One may therefore justifiably question what, if
anything, this solution has to do with black holes!

It is at this point that we should remember that the metric of the
spin-3 theory is not invariant under higher spin gauge transformations. In \GK\ it was therefore conjectured that there exists a spin-3
gauge transformation that will transform \cah\ into a black hole, and
evidence for this was presented at linear order in the charge.  In the next section we establish the validity of this conjecture to all orders, finding the explicit gauge transformation needed to take the wormhole into a black hole, along with the explicit black hole metric. This  metric is completely smooth, as is the
accompanying spin-3 field, and has AdS$_3$ asymptotics.
Furthermore, at least within our ansatz, there is a unique such black hole solution, and its smoothness requires that the holonomy conditions \cag\ be obeyed.

\newsec{Gauge transforming the wormhole into a black hole}

In this section we describe the solution to the following problem.
We start from the wormhole solution reviewed in the last section:
\eqn\caa{\eqalign{ \Ac&= \Big(e^\rho L_1 -{2\pi \over k}\Lc  e^{-\rho}L_{-1} -{\pi \over 2k} \Wc e^{-2\rho} W_{-2} \Big)dx^+  \cr
& \quad +\mu \Big(e^{2\rho} W_2 -{4\pi \Lc \over k}W_0 +{4\pi^2 \Lc^2 \over k^2} e^{-2\rho} W_{-2} +{4\pi \Wc \over k}e^{-\rho} L_{-1}\Big)dx^-  +L_0 d\rho  \cr
 \Acb&= -\Big(e^\rho L_{-1} -{2\pi \over k}\Lc e^{-\rho} L_{1} +{\pi \over 2k} \Wc e^{-2\rho} W_{2} \Big)dx^- \cr
 & \quad+ \mu \Big( e^{2\rho}  W_{-2} -{4\pi \Lc \over k}W_0 +{4\pi^2 \Lc^2 \over k^2} e^{-2\rho} W_{2} -{4\pi \Wc \over k}e^{-\rho} L_{1}\Big) dx^+ -L_0 d\rho  }}
We then consider new connections related to these by SL(3,R) gauge transformations:
\eqn\cab{\eqalign{ A &= g^{-1}(\rho) \Ac(\rho) g(\rho)+ g^{-1}(\rho) dg(\rho) \cr
 \Ab &= g(\rho) \Acb(\rho) g^{-1}(\rho)- dg (\rho) g^{-1}(\rho)  }}
with $g(\rho) \in ~ $SL(3,R).  The relative gauge transformation for
$\Ab$ versus $A$ is taken to maintain a non-rotating ansatz.  The
metric and spin-3 field corresponding to $(A,\Ab)$ will take the form
\eqn\cac{\eqalign{ ds^2  &= g_{\rho\rho}(\rho) d\rho^2  +g_{tt}(\rho) dt^2 + g_{\phi\phi}(\rho)d\phi^2 \cr
\varphi_{\alpha\beta\gamma} dx^\alpha dx^\beta dx^\gamma & = \varphi_{\phi \rho\rho}(\rho) d\phi d\rho^2 + \varphi_{\phi tt}(\rho) d\phi dt^2 + \varphi_{\phi\phi\phi}(\rho) d\phi^3 }}
We demand that this solution describe a smooth black hole with event horizon at $\rho=\rho_+$, or at $r=0$ with
\eqn\cada{r = \rho -\rho_+}
Assuming that $g_{rr}(0) >0$, as will be the case,
this first of all  requires $g_{tt}(0) = g_{tt}'(0)=0$ and $g_{\phi\phi}(0) >0$, so that after rotating to imaginary time the metric expanded around $r=0$ will look locally like $R^2 \times S^1$:
\eqn\cada{ ds^2 \approx g_{rr}(0) dr^2 - {1\over 2}g_{tt}''(0) r^2 dt_E^2 +g_{\phi\phi}(0)d\phi^2 }
In order for the metric to avoid a conical singularity at $r=0$  we need to identify $t_E \cong t_E + \beta$ with
\eqn\cae{ \beta = 2\pi \sqrt{ 2g_{rr}(0) \over -g_{tt}''(0)}}
Having done so, we can switch to Cartesian coordinates near $r=0$ and the metric will be smooth.

The same smoothness considerations apply to the spin-3 field.
Noticing the parallel structure, we see that we should demand
$\varphi_{\phi tt}(0)  = \varphi_{\phi tt}'(0)=0$, and
\eqn\caf{ \beta = 2\pi \sqrt{ 2\varphi_{\phi rr}(0) \over -\varphi_{\phi tt}''(0)}}
with the same $\beta$ as in \cae.

There is still one more condition to impose to ensure that the solution is completely smooth at the horizon.   If we work in Cartesian coordinates $(x,y)$ around $r=0$, we should demand that all functions
are infinitely differentiable with respect to both $x$ and $y$.  If this is not the case then some curvature invariant (or spin-3 quantity) involving covariant derivatives will diverge.   Given the rotational symmetry, this condition implies that the series expansion of all functions should only involve non-negative even powers of $r$.   We impose this by demanding that all functions be smooth at the horizon, and even under reflection about the horizon:
\eqn\cagg{\eqalign{ g_{rr}(-r)=g_{rr}(r)~,\quad  g_{tt}(-r)&=g_{tt}(r)~,\quad  g_{\phi\phi}(-r)=g_{\phi\phi}(r) \cr
\varphi_{\phi rr}(-r)=\varphi_{\phi rr}(r)~,\quad
\varphi_{\phi tt}(-r)&=\varphi_{\phi tt}(r)~,\quad
\varphi_{\phi \phi\phi}(-r)=\varphi_{\phi \phi\phi}(r) }}

We now summarize the solution to this problem.  More details are
provided in appendix D.   The symmetry conditions \cagg\ can be enforced by demanding
\eqn\cah{\eqalign{ e_t(-r) &= -h(r)^{-1}e_t(r)h(r)\cr
e_{\phi}(-r) &= h(r)^{-1}e_{\phi}(r)h(r)\cr
e_{r}(-r) &= h(r)^{-1}e_{r}(r)h(r)\cr}}
with $h(r) \in~ $SL(3,R), and similar conditions on the spin-connection. The BTZ solution has $h(r)={\bf1}$, so we can think of these conditions as a ``twist'' of the BTZ vielbein  reflection symmetries. In addition, $h(0)={\bf1}$, implying that $e_t(0)=0$, a feature of the BTZ solution that persists in the spin-3 case.\foot{Moreover, it is both surprising and convenient that the location of the horizon $r=0$ turns out to be at $\rho=\rho_+$, with $\rho_+$ given by the {\it same} expression as for BTZ: $e^{2\rho_+ } = {2\pi \Lc \over k}$.}

To gain some insight into the form of $g(r)$ and $h(r)$ we can start with the BTZ solution and then carry out the gauge transformation perturbatively in the charge.  These considerations lead us to the ansatz
\eqn\cai{\eqalign{  g(r) &= e^{F(r)(W_1-W_{-1})+G(r)L_0} \cr           h(r)&=  e^{H(r)(W_1+W_{-1})} }}
for some functions $F, G$ and $H$.   Perturbation theory suggests that this ansatz gives the unique solution to our problem, although we have not proven this.  On the other hand, having assumed the ansatz \cai\ the remaining analysis definitely has a unique solution.

Even after assuming this ansatz, finding a solution that satisfies all the smoothness conditions involves a surprisingly large amount of complicated algebra requiring extensive
use of Mathematica and Maple.  Some details are provided in the appendix.  As we have already mentioned, a crucial point is that the
solution to our problem requires that the holonomy conditions \cag\ are obeyed; equivalently, we can derive the holonomy conditions by requiring the existence of a smooth black hole solution.

Here we just present the final form for the transformed metric.
It will be convenient to define dimensionless versions of the charge and chemical potential:
\eqn\caj{\zeta = \sqrt{k \over 32 \pi \Lc^3}\Wc~,\quad \gamma = \sqrt{2\pi \Lc  \over k}\mu  }
as well as a parameter $C$ defined as
\eqn\cak{ \zeta = {C -1 \over C^{3/2} } }
The metric takes the form \cac\ with
\eqn\cala{\eqalign{ g_{rr} &= {(C -2)(C-3) \over \left(C -2-\cosh^2(r)\right)^2} \cr
 g_{tt} &=-\left({8\pi \Lc\over k}\right)\left(  {C -3 \over C^2} \right)  {\big(a_t+b_t \cosh^2(r)\big)\sinh^2(r) \over  \left(C -2-\cosh^2(r)\right)^2} \cr
 g_{\phi\phi} &= \left({8\pi \Lc\over k}\right)\left(  {C -3 \over C^2} \right)  {\big(a_\phi+b_\phi \cosh^2(r)\big)\sinh^2(r) \over  \left(C -2-\cosh^2(r)\right)^2}+\left({8\pi \Lc\over k}\right)(1+{16\over 3} \gamma^2+12 \gamma\zeta )  }}
The coefficients $a_{t,\phi}$ and $b_{t,\phi}$ are functions of $\gamma$ and $C$, and are displayed in appendix C along with the spin-3 field.

With these results in hand, we demand a smooth horizon via \cae\ and \caf. Using the definition \cak, the resulting equations can be written as
\eqn\cam{\eqalign{1728\gamma^3\zeta^2-(432\gamma^2+27)\zeta-128\gamma^3
+72\gamma&=0 \cr
\left( 1+ {16 \over 3} \gamma^2   -12 \gamma \zeta\right) \Lc -{\pi k \over 2 \beta^2}&=0} }
These are precisely the holonomy conditions \cag, merely rewritten in the $(\zeta,\gamma)$ variables!

\subsec{Limits and asymptotics}
Solution of equations \cam\ for the charge $\zeta$ and inverse temperature $\beta$ yields
\eqn\can{\eqalign{ \zeta & = {1+16\gamma^2 -\left(1-{16\over 3}\gamma^2\right)\sqrt{1+{128\over 3}\gamma^2} \over 128 \gamma^3}  \cr
\beta &= {   \sqrt{\pi k\over 2\Lc } \over \sqrt{ 1+ {16 \over 3} \gamma^2   -12 \gamma \zeta}}    } }
We have chosen the branch of $\zeta$ consistent with recovery of the BTZ solution in the $\gamma \rar 0$ limit, namely the one with a power series expansion in positive (odd) powers of $\gamma$.

The uncharged BTZ limit corresponds to taking $\zeta, \gamma \rt 0$,
and $C\rt \infty$.  On the other hand, the maximal value of
$(\zeta,\gamma)$ obtained from \can\ is
\eqn\cao{ \zeta_{max} = \sqrt{4\over 27}~,\quad \gamma_{max} = \sqrt{3\over 16}}
and this corresponds to $C=3$.   Therefore, we can take $C$ to lie in the range $3 \leq C < \infty$.

The extremal lower bound can also be seen directly on the level of the metric \cala, which degenerates at $C=3$. This is a virtue of the manifestly smooth horizon of the black hole, in contrast to the wormhole geometry which has no limiting form at this value of the charge, and hence one must resort to the holonomy to find it instead.

The metric coefficients diverge at $r=\rs$, where
\eqn\cap{ \cosh^2(r_\star) = C-2 }
The leading behavior of the metric near $r_\star$ is
\eqn\caq{ ds^2 \approx {1\over 4} {dr^2 \over (r_\star -r)^2} +\left({2\pi \Lc\over k}\right)\left( { C-3 \over C^2 (C-2)}\right){-[a_t + b_t(C-2)]dt^2+[a_\phi + b_\phi(C-2)]d\phi^2 \over (r_\star-r)^2}   }
This gives AdS$_3$ with radius $1/2$, and thus our transformed black hole solutions are asymptotically AdS$_3$.   In addition, the spin-3 field expressed in an orthonormal basis
goes to zero at $r_\star$.

As $C$ approaches $3$ from above we see from \cap\ that $r_\star \rt 0$.
Nonetheless we can extract the extremal asymptotics by scaling the coordinates as we take $C\rt 3$. This has the effect of separating $\rs$ from the horizon and stretching the region in between.

Expanding around extremality by defining
\eqn\car{\gamma= \gamma_{ext}-\delta}
we define asymptotic coordinates
\eqn\cas{r=\sqrt{\delta}\tilde{r}~,\quad t={\tilde{t}\over\delta}}
Expanding all quantities and taking the $\delta\rar0$ limit, one finds a metric
\eqn\cat{ds^2 \approx {\rts^2\over(\rts^2-\rti^2)^2}d\rti^2-{512\pi\ml\over9k}{\rts^2\over(\rts^2-\rti^2)^2}\rti^2d\tilde{t}^2+\left({32\pi\ml\over k}+{96\pi\ml\over k}{\rts^2\over(\rts^2-\rti^2)^2}\rti^2\right)d\phi^2}
where $\rts=\sqrt{8\sqrt{3}\over3}$. This metric has the same AdS$_3$ asymptotics as the nonextremal metric \caq, which becomes evident upon defining a Fefferman-Graham coordinate $\tilde{r} = \rts\tanh(\eta)$.

The coordinate $r$ appearing in \cala\ only covers the region outside the event horizon; it is analogous to $\sqrt{r-2M}$ for the Schwarzschild solution.  The region inside the horizon is obtained by continuing $r$ to pure imaginary values.

\subsec{Black hole entropy}
In \GK, the entropy of the black hole was found  to be
\eqn\apr{S = 4\pi\sqrt{2\pi k\ml} f(y)}
where
\eqn\aps{f(y)= \cos\theta \, , \quad \theta = {1\over 6}\arctan\left({\sqrt{y(2-y)}\over 1-y}\right)\, , \quad 0 \leq \theta \leq {\pi\over 6}}
and $y={27\over 2}\zeta^2$. (We have specialized to the static case.) This was originally found by solving a first-order nonlinear ODE, derived by combining the definition of the partition function with the holonomy conditions.

The function $f(y)$ takes a pleasantly simple form upon plugging in for $\zeta$ as a function of $C$, using \cak. This step yields
\eqn\apss{\theta = {1\over 6}\arctan\left(\Lambda(C)\sqrt{1-{3\over 4C}}\right)}
with
\eqn\apsss{\Lambda(C) \equiv {6\sqrt{3C}(C-1)(C-3)\over (2C-3)(C^2-12C+9)}}
Surprisingly, taking the cosine of this angle yields the simple expression
\eqn\apssss{f(y)=\sqrt{1-{3\over 4C}}}
As with other quantities in our analysis, we see that the entropy is most simply expressed in terms of $C$. The extremal and zero charge limits are recovered upon inspection.

It is of course natural to wonder if the black hole entropy can be expressed in terms of a geometrical property of the horizon.  There is of course no reason to expect that the Bekenstein-Hawking area law holds, since the spin-3 field is nonzero at the horizon, and indeed one easily checks that $S \neq A/4G$.   This is related to one of the primary challenges inherent in higher spin gravity: the enlarged gauge invariance renders familiar geometric quantities, such as the horizon area, non-invariant under higher spin gauge transformations.  Perhaps there exists a higher spin version of the Wald entropy formula \WaldNT\ that is fully gauge invariant.

\newsec{Discussion}

Let us close with some open issues. We address the black hole, the RG flow, and generalization to other higher spin theories in turn.

Our main result was showing explicitly how to gauge transform between the wormhole and the black hole.  The overall logical structure is very tight: the existence of the gauge transformation is contingent upon the holonomy conditions being satisfied, and there then exists a unique gauge transformation and smooth black hole metric.\foot{Strictly speaking, these statements are subject to the qualifications noted below \cai.}  The wormhole and black hole have dramatically different causal structures, and so it is of course conceptually interesting that they can be related by a higher spin gauge transformation.

 This situation has no analog in ordinary gravity.  By adding in matter to probe the solution, one can map out the light cones and determine the causal structure in a unique fashion.
But  in our spin-3 theory it is not possible to simply throw in a minimally coupled scalar field to probe the solution, as there is no obvious way to do
so that is compatible with the full higher spin gauge invariance.  In this respect, the situation is analogous to string theory, where it is also not possible to add in matter arbitrarily.  However, it is known how to couple in propagating scalar fields to the large $N$ limit of these higher spin theories \refs{\ProkushkinBQ,\ProkushkinVN}\foot{Using this formalism massless fields in the background of the BTZ black hole were studied in \DidenkoZD.}, and such scalars play an important role in the conjecture of \GaberdielPZ.  It therefore seems
possible to compute AdS/CFT correlators between the two asymptotic boundaries of the wormhole/blackhole.  One could then determine whether or not these two boundaries are causally connected, by seeing whether such correlators exhibit lightcone singularities.

We have already noted that it would be very useful to have a higher spin version of Wald's  entropy formula at our disposal to gain a better
geometric understanding of the theory, to the extent that this is possible.
Our entropy formula  \apssss\ takes a surprisingly simple form (as compared to the  algebra leading to it), and is perhaps suggestive of a geometrical interpretation.  Also, the twisted vielbein reflection conditions \cah\
deserve to be better understood.

While the existence of the black hole gauge is extremely useful for
conceptual and interpretational purposes, it is  likely to be the case that the wormhole gauge is more convenient for doing computations. In the wormhole gauge we know how to read off the  charges and symmetries from the asymptotic form of the connection.  In fact, this can be done in either of two ways: by viewing the solution in terms of the $W_3$ CFT deformed by an irrelevant
spin-3 operator, or in terms of the $W_3^{(2)}$ CFT deformed by a relevant spin-3/2 operator. On the other hand, in the black hole gauge, the gauge field near
$r=\rs$ does not  take a form in which we know how to identify the CFT data,   cf. \wtta.  It would be convenient and perhaps enlightening if we could understand the black hole asymptotics better. Such an analysis would likely need to involve the subleading terms near the AdS$_3$ boundary at $r=\rs$.

The existence of multiple AdS$_3$ vacua in this theory, both of which have vanishing spin-3 field, is another intriguing and novel feature due to the inclusion of higher spin.
We were led to a simple RG flow solution between these vacua triggered by a spin-3/2, Lorentz symmetry-breaking CFT operator. We would like to improve our understanding of the behavior of the central charge under the flow, and in particular why it increases towards the IR. As noted earlier, the Lorentz symmetry-breaking nature of this RG flow places it outside the assumptions of the $c$-theorem; maybe another quantity can be constructed which monotonically decreases along RG flows from UV to IR.

Perhaps most conceptually interesting is the extension of these ideas to bulk theories with larger gauge groups. Let us consider SL(N,R) $\times$ SL(N,R) for now. To each embedding of SL(2,R) in SL(N,R) is associated an AdS$_3$ vacuum with asymptotic symmetry given by the $W$-algebra obtained by classical Drinfeld-Sokolov reduction. Modulo issues of interpretation described above, the central charge of the Virasoro algebra coming from the principal embedding will always be the largest. That of the other vacua will have an inverse relation to the index of the SL(2,R)  embedding used to construct the vacuum. Some relevant group theoretic details are presented in appendix E.

It is easy to write these solutions down in the Chern-Simons language once one has the explicit SL(2,R) embedding: simply take ansatz \aih\ and replace $\Wh_2$ and $\Lh_0$ by the generators of the new SL(2,R) embedding. It is equally straightforward to construct RG flows between these vacua, and hence between CFTs with different symmetries, by altering \aja\ in the same manner. The set of allowed RG flows depends on the details of the Lie algebra, but one can see that in general the vacuum with largest (smallest) AdS radius has no relevant (irrelevant) operators, and so is IR (UV) stable.

If one takes the $N \rar \infty$ limit, one finds a discretuum of AdS$_3$ vacua, each with some asymptotic $W$-algebra with an infinite number of primary fields. The distribution of central charges in this limit is subtle, as one must now take care to define the limit of large $k$ and large $N$ appropriately. These sorts of questions have been addressed in similar contexts in recent higher spin literature, e.g. \refs{\HenneauxXG, \GaberdielPZ, \CastroCE}. This is a fascinating implication of the large $N$ limit that deserves to be better understood, and we leave further investigation for future work.

\vskip .1in

\noindent
{ \bf Acknowledgments}

\vskip .2cm

 This work was supported in part by NSF grant PHY-07-57702. We are grateful to  M. Ba\~{n}ados, A. Castro, A. Maloney and D. Marolf  for  useful conversations and correspondence.

\appendix{A}{SL(3,R) generators}

As in  \CampoleoniZQ\ with $\sigma =-1$, we use the following basis of SL$(3,R)$ generators
\eqn\za{ \eqalign{ L_1 & = \left(\matrix{0&0&0 \cr 1&0&0 \cr 0&1&0}\right),\quad L_0= \left(\matrix{1&0&0 \cr 0&0&0 \cr 0&0&-1}\right),\quad  L_{-1} = \left(\matrix{0&-2&0 \cr 0&0&-2 \cr 0&0&0}\right)\cr & \cr
W_2 &=  \left(\matrix{0&0&0 \cr 0&0&0 \cr 2&0&0}\right),\quad W_1 =  \left(\matrix{0&0&0 \cr 1&0&0 \cr 0&-1&0}\right),\quad  W_0 = {2\over 3} \left(\matrix{1&0&0 \cr 0&-2&0 \cr 0&0&1}\right)\cr & \cr
W_{-1} &=  \left(\matrix{0&-2&0 \cr 0&0&2 \cr 0&0&0}\right),\quad W_{-2} =  \left(\matrix{0&0&8 \cr 0&0&0 \cr 0&0&0}\right) }}
The generators obey the following commutation relations
\eqn\zb{\eqalign{ [L_i ,L_j] &= (i-j)L_{i+j} \cr
[L_i, W_m] &= (2i-m)W_{i+m} \cr
[W_m,W_n] &= -{1 \over 3}(m-n)(2m^2+2n^2-mn-8)L_{m+n} }}
and trace relations
\eqn\zc{ \eqalign{ \Tr ( L_0 L_0) &= 2~,\quad   \Tr ( L_1 L_{-1} ) = -4 \cr   \Tr ( W_0 W_0 )  &= {8 \over 3}~,\quad \Tr ( W_1 W_{-1} )  = -4 ~,\quad \Tr ( W_2 W_{-2} )  = 16   }}
All other traces involving a product two generators vanish.

\appendix{B}{Stress tensor correlator in RG flow solution}

In this appendix we compute the AdS/CFT two-point correlator of the
UV stress tensor $T_{UV}$ in the background of the RG flow solution \aja.  To compute the correlator we need to turn on the source conjugate to $T_{UV}$ and then compute the linearized response.  The coefficient relating $T_{UV}$ to the source is the two-point function. In this computation we will not pay attention to overall normalization factors.

In studying  linearized fluctuations around the RG solution,
the following ansatz turns out to be appropriate
\eqn\yaa{\eqalign{ A & = \left( \mu e^{2\rho} W_2 +\lambda e^{\rho}L_1 + l_0 L_0 +  h_1 e^{-\rho}L_{-1} + h_2e^{-2\rho} W_{-2} \right) dx^+ \cr
& +\left( W_2e^{2\rho} + J_{UV} W_0 + G^{(1)}_{UV}e^{-\rho}L_{-1} + G^{(2)}_{UV}e^{-\rho} W_{-1} + T_{UV} e^{-2\rho}W_{-2}\right) dx^- +L_0 d\rho \cr
\Ab &  =   e^{2\rho} W_{-2}dx^+ -\lambda e^{\rho}L_{-1}dx^-  -L_0 d\rho}}
Here $\lambda$ is constant, while all other coefficients are arbitrary functions of $x^\pm$.   The leading large $\rho$ behavior of the metric derived from this connection is
\eqn\yab{ ds^2 \approx d\rho^2  - 4 e^{4\rho} dx^+ dx^- -4 e^{4\rho} \mu (dx^+)^2 }
From this we see that $\mu$ acts as a source for $T_{++}=T_{UV}$, and
so we need to work out the relation between $T_{UV}$ and $\mu$.
Solving the linearized flatness conditions, we find
\eqn\yac{\eqalign{ G^{(2)}_{UV} & = -{1\over 3\lambda} \p_+ J_{UV} \cr
l_0 &= -{1\over 2} \p_- \mu \cr
h_1& = {\lambda \over 2}J_{UV} \cr
h_2 & = -{1\over 32} \p_-^2 \mu -{\lambda \over 8} G^{(1)}_{UV}  }}
along with
\eqn\yad{\eqalign{ &\p_+ G^{(1)}_{UV} = {\lambda \over 2} \p_- J_{UV} \cr
 &8 \p_+^3 J_{UV} +12 \lambda^3 \p_- G^{(1)}_{UV} = -3\lambda^2 \p_-^3 \mu \cr
& T_{UV} = {1\over 12\lambda^2} \p_+^2 J_{UV} }}
We solve \yad\ in momentum space, assuming dependence $e^{ip_+ x^++i p_- x^-}$, which gives
\eqn\yae{\eqalign{ T_{UV} &= -{1\over 24 } {p_+^3 p_- \over \lambda^4- {4\over 3}{p_+^4 \over p_-^2 }}\mu   } }
which implies the result quoted in \ajf.

\appendix{C}{Black hole and spin-3 field parameters}
\subsec{Metric and spin-3 field in black hole gauge}
The coefficients $a_{t,\phi}$ and $b_{t,\phi}$ in the black hole metric \cala\ are as follows:
\eqn\apa{\eqalign{a_t &= (C-1)^2\left(4\gamma-\sqrt{C}\right)^2\cr
a_{\phi} &= (C-1)^2\left(4\gamma+\sqrt{C}\right)^2\cr
b_t &= 16\gamma^2(C-2) (C^2-2C+2)-8\gamma\sqrt{C} (2C^2-6C+5)+C(3C-4)\cr
b_{\phi} &= 16\gamma^2(C-2) (C^2-2C+2)+8\gamma\sqrt{C} (2C^2-6C+5)+C(3C-4)\cr}}
The $t$ and $\phi$ coefficients are related by flipping the sign of $\gamma$, though this is not a bonafide sign flip of the charge, under which $C$ would also transform.

The spin-3 field has components
\eqn\apb{\eqalign{\vphi_{\phi rr} &= {2\over 3}\sqrt{{2\pi\ml\over k}}{(C-3)(4\gamma(C^2-5C+3)-3\sqrt{C})\over C(C-2-\cosh^2(r))^2}\cr
\vphi_{\phi tt} &= -{16\sqrt{2}\over 3}\left({\pi\ml\over k}\right)^{3/2}\left({C-3\over C^3}\right){(a_{t,3}+b_{t,3}\cosh^2(r))\sinh^2(r)\over(C-2-\cosh^2(r))^2}\cr
\vphi_{\phi\phi\phi} &=16\sqrt{2}\left({\pi\ml\over k}\right)^{3/2}\left({C-3\over C^3}\right){(a_{\phi,3}+b_{\phi,3}\cosh^2(r))\sinh^2(r)\over(C-2-\cosh^2(r))^2}+ \vphi_{\phi\phi\phi}(0)\cr}}
where

\eqn\apc{\eqalign{a_{t,3} &=  a_t\cdot\left(4\gamma(3-2C)-3\sqrt{C}\right)\cr
a_{\phi,3} &= a_{\phi}\cdot\left(4\gamma(3-2C)-3\sqrt{C}\right)\cr
b_{t,3} &= 64\left(C(C(C^2-10C+30)-37)+15\right)\gamma^3\cr&-16\sqrt{C}(C(10C^2-57C+88)-36)\gamma^2\cr&+4C(C(9C^2-42C+62)-36)\gamma\cr& -3C^{3/2}(2C^2-6C+5)\cr
b_{\phi,3} &=  64\left(C(C(C^2-10C+30)-37)+15\right)\gamma^3\cr&-16\sqrt{C}(C(2C^2+3C-16)+12)\gamma^2\cr&+4C(C(-3C^2+6C-10)+12)\gamma\cr& -3C^{3/2}(2C^2-6C+5)\cr}}
and
\eqn\apcc{\eqalign{
\vphi_{\phi\phi\phi}(0) &= {16\sqrt{2}\over 9C^3}\left({\pi\ml\over k}\right)^{3/2}\left(4\gamma(2C-3)+3\sqrt{C}\right)\cdot\cr&\quad\left(16\gamma^2(C^2-12C+9)+12\gamma\sqrt{C}(3-5C)-9C(C-1)   \right)\cr}}
\subsec{Horizon holonomy}
For completeness, we provide the holonomy around the $\phi$ circle, which can be viewed as another piece of gauge-invariant information characterizing the effect of spin-3 charge. The holonomy matrix is just $A_{\phi}$ itself. We again work with the trace squared and determinant to obtain
\eqn\apcccc{\eqalign{\det(A_{\phi}) &= -16\left({2\pi\ml\over k}\right)^{3/2}(1+16\gamma^2)\zeta\cr \Tr(A_{\phi}^2) &= \left({16\pi \Lc\over k}\right)(1+{16\over 3} \gamma^2+12 \gamma\zeta ) \cr}}
The holonomy for the barred gauge field is
\eqn\apccccc{\det(\bar{A}_{\phi}) = -\det(A_{\phi})~, \quad \Tr(\bar{A}_{\phi}^2) = \Tr(A_{\phi}^2)}
Note from \cala\ that  $\Tr (A_\phi^2)$ is directly related to the area of the event horizon, since $g_{\phi\phi}(0) = {1\over 2} \Tr (A_\phi^2)$.

\appendix{D}{From wormhole to black hole}
We begin by recalling the vielbein reflection equations \cah:
\eqn\apd{\eqalign{ e_t(-r) &= -h(r)^{-1}e_t(r)h(r)\cr
e_{\phi}(-r) &= h(r)^{-1}e_{\phi}(r)h(r)\cr
e_{r}(-r) &= h(r)^{-1}e_{r}(r)h(r)\cr}}
where $h(r) \in SL(3,R)$. We obtained these by solving for the horizon geometry perturbatively in the charge, noticing that $e_t(0)=0$, and demanding reflection symmetry of the metric and spin-3 field as one moves away from the horizon. Consideration of the spin-connection, for which only $\omega_{\phi}(0)=0$, allows us to convert these to statements about the gauge fields which are simpler to work with:
\eqn\ape{\eqalign{A_+(-r) &= h^{-1}(r) \Ab_-(r) h(r) \cr \Ab_+(-r) &= h^{-1}(r) A_-(r) h(r)\cr   A_r(-r) -\Ab_r(-r) &=  h^{-1}(r) \big[  A_r(r) - \Ab_r(r)\big] h(r) \cr
 A_r(-r) + \Ab_r(-r) &= \alpha(r) h^{-1}(r) \big[  A_r(r) + \Ab_r(r)\big] h(r) }}
$\alpha(r)$ is some function of $r$ and the charge which will not be needed.

Recalling that the gauge field $A$ is related to the wormhole gauge field $\cal{A}$ by \cab, our goal is to solve  equations \ape\ for $g(r)$ while solving for $h(r)$ along the way. We will solve the first of these equations, after which we find that the remaining three are satisfied automatically.

As stated in the text, perturbation theory indicates that $g(r)$ and $h(r)$ take the following simple forms:
\eqn\appe{\eqalign{  g(r) &= e^{F(r)(W_1-W_{-1})+G(r)L_0} \cr           h(r)&=  e^{H(r)(W_1+W_{-1})} }}
This ansatz gives a metric and spin-3 field which respect the symmetries of the static BTZ solution around which we perturb. $F$ and $H$ are odd in $\gamma$, and $G$ is even. In addition, $H$ is odd under reflection through the horizon, consistent with
\eqn\apf{h^{-1}(-r) = h(r)\equiv h}
which is implied by \apd. Notice that the problem is now highly overconstrained: we are solving for three functions $(F,G,H)$, but we have four $3 \times 3$ matrix equations to solve.

Let us rewrite the first of equations \ape\ in terms of $\Ac$:
\eqn\apg{\Ac_+(-r)=M^{-1}\Acb_-(r)M~, \quad M=g^{-1}(r)h(r)g^{-1}(r)}
There are five independent components of this matrix equation, which we solve directly. Expanded in generators, $M$ includes pieces proportional to each element of SL(3,R) and the identity, making a mess of algebra. One is aided by defining redundant variables, solving for them, and reinserting these definitions to solve for $(F,G,H)$. To that end, we define the variables $(X,Y)$ as
\eqn\aph{Y=-{\sqrt{4F^2+G^2}\over G}~, \quad X = e^{-GY}}
These combinations are ubiquitous in the explicit form of these equations.

After a display of brute force, one can reduce these five equations to the following three:
\eqn\api{\eqalign{\zeta &= {Y^2-(1+\cosh^2(r))\over(Y^2-1)^{3/2}}\cosh (r)\cr
X &= \sqrt{Y-1\over Y+1}\sqrt{Y+1+\cosh^2(r)\over Y-(1+\cosh^2(r))}\cr
\tan H&=-{\sinh(r)\cosh(r)\over \sqrt{Y^2-(1+\cosh^2(r))^2}}}}
The sign of $H$ correlates with the convention $\mu>0$.

One can solve the first equation by taking
\eqn\apj{Y^2 = 1+C\cosh^2(r)}
where $C$ is defined by
\eqn\apl{\zeta = {C-1\over C^{3/2}}}
as in \cak. The final expression for $X$, and then for $(F,G,H)$, can be written most compactly as
\eqn\apm{\eqalign{X &= \sqrt{{C+Y-1\over C-Y-1}}\cr
G &= -{1\over Y} \log(X)\cr
{F\over G} &= {\sqrt{C}\over 2}\cosh (r)\cr
\tan H &=-{\sinh(r)\over\sqrt{C-2-\cosh^2(r)}}}}
Remarkably, this solves all of the reflection symmetry equations \ape.

Let us state some of the parameter ranges. Recalling the holonomy conditions, for example, we know that $3 \leq C < \infty$. This implies $|Y|\geq 2$, so
choosing the branch $Y>0$, we see that $X\geq 1$. This in turn implies that $(F,G)<0$.

We can clearly see the divergence that feeds down to the metric and spin-3 field. Using the definition of $Y$, we note that
\eqn\apn{\cosh^2(\rs) = C-2 \quad \Leftrightarrow \quad Y(\rs)=C-1}

In the zero charge limit, $C\rightarrow \infty$, $X\rt 1$, and $(F,G,H)\rt 0$ as we recover the BTZ black hole. In the extremal limit, $C\rt 3$, $X \rt \infty$, and $(F,G) \rt -\infty$. This explains the divergence of the metric and spin-3 field in the extremal limit: the gauge parameters are breaking down.
For convenience, we present the first few terms in a perturbative expansion of $F$ and $G$. With
\eqn\apo{\eqalign{F&= f_1\gamma+f_3\gamma^3+f_5\gamma^5+\ldots\cr
G&=g_2\gamma^2+g_4\gamma^4+g_6\gamma^6+\ldots\cr}}
we have
\eqn\app{\eqalign{
f_1&= -{4\over 3}\cosh(r)\cr
f_3&=-{128\over 81}\cosh(r)\left(2\cosh^2(r)-3\right)\cr
f_5&=-{8192\over3645}\cosh(r)\left(6\cosh^4(r)-15\cosh^2(r)+40\right)\cr}}
and
\eqn\apq{\eqalign{
g_2 &= -{64\over9}\cr
g_4&=-{4096\over243}\left(\cosh^2(r)-6\right)\cr
g_6&=-{65536\over10935}\left(12\cosh^4(r)-60\cosh^2(r)+395\right)\cr}}

\appendix{E}{Generalization to  SL(N,R)}
In this appendix  we will briefly discuss (mainly following \refs{\BaisBS,\ForgacsAC\deBoerIZ}) how one determines the asymptotic symmetry algebras for the SL(N,R) vacua.  As shown in  \DynkinUM\ the inequivalent SL(2,R) embeddings in SL(N,R)  are  uniquely determined by the branching of the fundamental representation of SL(N,R) into $n_i$ dimensional representations of SL(2,R). The branchings are given by the partitions $\{ n_1,n_2,\cdots , n_l\} $ of $N$.   From this one can determine how the adjoint, i.e. the algebra itself, decomposes into representations of the SL(2,R) algebra. In general there are representations of spin $s=0$ up to spin $s=N-1$. We denote the number of spin $s$ representations by $m_s$ (of which some can be zero).
For example the principal embedding is given by the partition $\{N\}$, hence the adjoint representation decomposes  as
\eqn\appda{m_1=1,m_2=1,\cdots ,m_{n-1}=1}
We denote the  SL(2,R) generators as $(\hat{L}_{+},\hat L_- ,\hat L_0)$ (corresponding to a spin s=1 multiplet) and the generators of spin $s_i$ as $(W^{(i)}_{-s_i}, W^{(i)}_{-s_i+1},\cdots ,W^{(i)}_{s_i})$.
The ansatz for the  SL(N,R) connection is a ``highest weight" gauge, where we associate a field with the  $W^{(i)}_{-s_i}$ for each spin $i$.
\eqn\appdb{A= e^{-\rho \hat L_0}  \Big( \hat L_1+ {\cal T} (x^+)\hat L_{-1}   +\sum_i  J^{(i)}(x^+)  W^{(i)}_{-s_i}\Big)  e^{\rho \hat L_0} dx^+ + \hat L_0 d\rho}
Here ${\cal T}$ is related to the stress energy tensor of the conformal algebra.
A general gauge transformation is given by
\eqn\appdc{\eqalign{\lambda &= e^{-\rho \hat L_0} \Big( \epsilon_1 \hat L_1+ \epsilon_0  \hat L_0 + \epsilon_{-1}\hat L_{-1}  + \sum_i \sum_{j=0}^{2 s_i}  \alpha^{(i)}_j  W^{(i)}_{-s_i+j} \Big) e^{\rho \hat L_0}  }}

By using the fact that the $W^{(i)}_{s}$ transform in spin $s_i$ representations of the SL(2,R) and following the general strategy of considering gauge transformations which preserve the gauge choice \appdc, one can establish the following facts:

\medskip

First, for each field $J^{(i)}$ with spin zero  (i.e. $s_i=0$) to transform like a conformal primary, the relation of ${\cal T}$ to the stress tensor and the transformation parameter have  to be modified. Temporarily referring only to the $\{J^{(i)},\alpha_0^{(i)}\}$ of the spin-0 representations, we must make the redefinitions
\eqn\appdd{T = {\cal T} + \sum_i  {1\over 2} (J^{(i)})^2, \quad \quad \alpha^{(i)}_{0} = \tilde \alpha^{(i)}_0+ \epsilon_1 J^{(i)} }
With this improvement $J^{(i)}$ is associated with a weight one primary, and $T$ is the stress tensor up to constant rescaling.
Second,  the fields $J^{(k)}$  with spin $s_k>0$ then all transform like conformal primaries of weight $s_k+1$.

For the principal embedding,  \appda\ implies that one has  a (quasi) primary of weight 2, i.e. the stress tensor, and conformal primaries of weight $3,4,\cdots, n$. This is indeed the field content of the $W_n$ algebra and the classical $W_n$ algebra can be obtained this way.

Other SL(2,R) embeddings will lead to different $W$ algebras. For example, for SL(3,R)  the only other partition is given by $\{2,1\}$, and  \appda\ becomes
\eqn\appde{m_0 =1, \; m_{1/2} =2, \; m_1=1}
This gives one weight 1, two weight 3/2 and one weight 2 primary of the $W_3^{(2)}$ algebra,  as was established in more detail in the main part of the paper.

The number of embeddings quickly grows with larger $N$ and we will not discuss these cases here.

\listrefs
\end